%% file: ms.tex
\documentclass[12pt,preprint]{emulateapj}

\usepackage{lscape}

\slugcomment{Received 2008 October 27; accepted by ApJ 2009 January 29}

\shorttitle{Disk-Braking in Young Stars}
\shortauthors{Nguyen et al.}

\begin{document}


\title{Disk-Braking in Young Stars: Probing Rotation in Chamaeleon~I and Taurus-Auriga}


\author{Duy Cuong Nguyen\altaffilmark{1}, Ray Jayawardhana\altaffilmark{1}, Marten H. van Kerkwijk\altaffilmark{1}, Alexis Brandeker\altaffilmark{2}, Alexander Scholz\altaffilmark{3}, Ivana Damjanov\altaffilmark{1}}

\email{nguyen@astro.utoronto.ca}

\altaffiltext{1}{Department of Astronomy \& Astrophysics, University of Toronto, 50 St. George Street, Toronto, 
ON M5S 3H4, Canada}
\altaffiltext{2}{Department of Astronomy, Stockholm Observatory, SE-106 91 Stockholm, Sweden}
\altaffiltext{3}{SUPA, School of Physics \& Astronomy, University of St. Andrews, North Haugh, St. Andrews, 
KY16 9SS, United Kingdom}

\begin{abstract}
We present a comprehensive study of rotation, disk and accretion signatures for 144 T Tauri stars in the young ($\sim\!2$ Myr old) Chamaeleon~I and Taurus-Auriga star forming regions based on multi-epoch high-resolution optical spectra from the Magellan Clay 6.5 m telescope supplemented by mid-infrared photometry from the {\it Spitzer Space Telescope}. In contrast to previous studies in the Orion Nebula Cluster and NGC~2264, we do not see a clear signature of disk braking in Tau-Aur and Cha~I. We find that both accretors and non-accretors have similar distributions of $v~\!\sin~\!i$. This result could be due to different initial conditions, insufficient time for disk braking, or a significant age spread within the regions. The rotational velocities in both regions show a clear mass dependence, with F--K stars rotating on average about twice as fast as M stars, consistent with results reported for other clusters of similar age. Similarly, we find the upper envelope of the observed values of specific angular momentum $j$ varies as $M^{0.5}$ for our sample which spans a mass range of $\sim\!0.16\,M_{\sun}$ to $\sim\!3\,M_{\sun}$. This power law complements previous studies in Orion which estimated $j\!\propto\!M^{0.25}$ for $\lesssim\!2$\,Myr stars in the same mass regime, and a sharp decline in $j$ with decreasing mass for older stars ($\sim\!10$\,Myr) with $M\!<\!2\,M_{\sun}$. Furthermore, the overall specific angular momentum of this $\sim\!10$\,Myr population is five times lower than that of non-accretors in our sample, and implies a stellar braking mechanism other than disk braking could be at work. For a subsample of 67 objects with mid-infrared photometry, we examine the connection between accretion signatures and dusty disks: in the vast majority of cases (63/67), the two properties correlate well, which suggests that the timescale of gas accretion is similar to the lifetime of inner disks.
\end{abstract}


\section{Introduction}
\label{sec-Introduction}

One of the major outstanding issues in star formation theory is the regulation of angular momentum in young stars. The specific angular momentum of young single stars at $\sim\!1$\,Myr is about four orders of magnitude lower than in molecular cloud cores, from which the stars formed, indicating efficient rotational braking in the early phases of stellar evolution \citep{Bodenheimer:1995p3501}. In this context, a large number of studies have explored the connection between the presence of disks and rotation \citep{Herbst:2007p3967}.

Disk braking is defined here as a process that provides rotational braking based on magnetic coupling between the star and the disk. One possible theoretical scenario for disk braking is `disk-locking', originally proposed for T Tauri stars by \citet{Camenzind:1990p1102}, \citet{Koenigl:1991p224}, and \citet{Shu:1994p1082}. In that case, the magnetic connection between the star and the disk produces a torque onto the star, transfering angular momentum to the disk (presumably, from where it is eventually removed by, e.g., magnetically driven winds). An alternative scenario for disk braking is stellar winds powered by accretion, as recently modeled by \citet{Matt:2005p4028}. For a more detailed overview of the theoretical work on disk braking, see for example the review by \citet{Matt:2008p4022}.

If disk braking is at work, we expect to observe three kinds of stars: slow rotators with disks, slow rotators without disks, and fast rotators without disks. This distribution corresponds to the following evolutionary sequence: while stars are coupled to their disks, they will rotate slowly; once stars lose their disks, they will continue to rotate slowly for some time but gradually spin-up as they contract towards the main sequence, with some stars eventually becoming fast rotators. Thus, rapidly rotating stars with disks should not exist in an ideal disk braking scenario.

Observationally, the evidence for disk braking is confusing. Some photometric studies found a correlation between rotational properties and near-infrared color excess suggestive of disks \citep[e.g.][]{Edwards:1993p2744,Herbst:2002p993}, whereas others have not \citep[e.g.][]{Stassun:1999p1017,Makidon:2004p4179}. The photometric monitoring program of \cite{Lamm:2005p3547} observed disk braking in $\sim\!2$--$3$\,Myr NGC~2264, but with the effect less pronounced for low mass stars. Recent studies using Spitzer mid-infrared observations of the $\sim\!1$\,Myr Orion Nebula Cluster (ONC) and NGC~2264 support a disk-rotation connection: stars with longer rotation periods were found to be more likely than those with short periods to have IR excesses \citep{Rebull:2006p190,Cieza:2007p3508}. However, the mid-infrared study by \citet{Cieza:2006p3509} of IC 348 did not find the preferential distribution of rotation with disk presence.

While both near-infrared and mid-infrared signatures indicate the presence of a dusty disk, they do not prove the coupling between star and disk as required by the disk braking scenario. To demonstrate a direct link between the inner disk and the central star, a better diagnostic for disk braking may be ongoing accretion. For strongly active accretors, rotation periods are difficult to determine since period measurements rely on the presence of stable starspot regions; therefore, period samples may be biased towards weakly accreting stars. In some respects, it is advantageous to use projected rotational velocity $(v~\!\sin~\!i)$ instead of rotation periods.
 
A recent spectroscopic study of disk accretion in low-mass young stars by \citet{Jayawardhana:2006p1156} found evidence of a possible accretion-rotation connection in the $\eta$~Cha ($\sim\!6$\,Myr) and TWA ($\sim\!8$\,Myr) associations. All accretors in their sample of 41 stars were slow rotators, with $v~\!\sin~\!i \lesssim 20$~km~s$^{-1}$, whereas the non-accretors showed a large span in rotational velocities, up to $50$~km~s$^{-1}$. However, given the small number of accretors, they caution that those results should be checked with larger samples. A larger study of solar-like mass stars in NGC~2264 by \citet{Fallscheer:2006p3553} found disk braking signatures in using UV excess indicative of accretion. For a review of recent observational studies on rotation and angular momentum evolution of young stellar objects and brown dwarfs, see \citet{Herbst:2007p3967}. 

As part of a comprehensive, multi-epoch spectroscopic survey, we present here a study of rotation and disk-braking at ages of $\sim\!2$\,Myr in the star forming regions Taurus-Auriga and Chamaeleon~I. This study comprises 144 stars, which significantly enlarges the previously available sample of spectroscopic data in those two regions (see the summary by \citet{Rebull:2004p3590} for currently available rotational data). From the spectra, we extract $v~\!\sin~\!i$, and, as accretion indicators, the full width of H$\alpha$ at $10\%$ of the peak (hereafter, H$\alpha$ $10\%$ width) and \ion{Ca}{2} fluxes. We investigate the distribution of $v\sin i$, estimate angular momentum values, and test for the signature of disk braking.

\section{Target Selection and Observations}
\label{sec-Observations}

We used $572$ high-resolution optical spectra of $144$ members in the $\sim\!2$\,Myr old Chamaeleon~I and Taurus-Auriga star forming regions obtained with the echelle spectrograph MIKE \citep{Bernstein:2003p626} on the Magellan Clay 6.5 meter telescope at the Las Campanas Observatory, Chile. The data were collected on 15 nights during four observing runs between 2006 February and 2006 December. We complemented our optical spectra with infrared measurements from the InfraRed Array Camera \citep[IRAC;][]{Fazio:2004p4374} aboard the {\it Spitzer Space Telescope}. For Cha~I, we used the results of \citet{Damjanov:2007p146}, and for Tau-Aur we analyzed publicly available images obtained between 2004 September and 2007 March, using the methods described in detail by \citet{Damjanov:2007p146}. Our results are listed in tables \ref{tbl-SummaryCha} \& \ref{tbl-SummaryTau}.

Our sample consists of a magnitude-limited subset (R $\le 17.6$ for Cha~I; R $\le 13.4$ for Tau-Aur) of targets from \citet{Leinert:1993p655}, \citet{Ghez:1993p656}, \citet{Simon:1995p657}, \citet{Kohler:1998p658}, \citet{Briceno:2002p651}, and \citet{Luhman:2004p647, Luhman:2004p653}. To isolate the possible influence of binarity on disk braking in this study, we excluded from our sample unresolved wide binaries and double-lined spectroscopic binaries. Our targets span the spectral type range from F2 to M5 based on published classifications. In addition, we observed a sample of 25 slowly rotating velocity standard stars selected from the list of \citet{Nidever:2002p519}; these cover the same spectral range as our targets. We determined the spectral type for 13 targets without prior classification by fitting their spectra against those of the standard stars, and identifying the best fits.

MIKE is a slit-fed double echelle spectrograph with blue and red arms. For this study, we used only the red spectra, which cover the range of $4\,900$--$9\,300$\,\AA\, in 34 spectral orders. The $0.35''$ slit was used with no binning to obtain the highest possible spectral resolution, R $\sim\!60\,000$. The pixel scale was $0.14''$ pixel$^{-1}$ in the spatial direction, and approximately $0.024$\,\AA\,pixel$^{-1}$ at $6\,500$\,\AA\, in the spectral direction. In MIKE, the spatial direction of the projected slit is wavelength dependent, and not aligned with the CCD columns. To extract these slanted spectra, we used customized routines running in the ESO-MIDAS environment (described in detail in Brandeker et al., in preparation). Integration times were chosen such that we obtained signal-to-noise ratios (S/N)\,$>\!30$ per spectral resolution element at $6\,500$\,\AA; they typically ranged from $60$ to $1\,200$ seconds depending on seeing.

\section{Analysis}
\label{sec-Analysis}

\subsection{Accretion Signatures}
\label{sec-Accretion}

The H$\alpha$ equivalent width (EW) has long been used to distinguish accreting or classical T Tauri stars (EW $>\!10$\,\AA) from non-accreting or weak-line T Tauri stars (EW $<\!10$\,\AA). In accretors, in the context of the magnetospheric accretion scenario, the H$\alpha$ emission arises largely from the gas falling in from the inner disk edge onto the star. In non-accretors, chromospheric activity is the main source of H$\alpha$ emission, and is thus generally weaker. The H$\alpha$ profiles of accretors also tend to be much broader than those of non-accretors due to the high velocity of the infalling gas and Stark broadening. (The latter is expected to be important in H$\alpha$, since the line optical depths are high; see \citet{Muzerolle:2001p2783} for further discussion.) Asymmetry in the H$\alpha$ profile of accretors is also commonly observed as a result of viewing geometry, absorption by a wind component, or both.

Since the H$\alpha$ EW depends on the spectral type, \citet{White:2003p1049} proposed to use as a more robust accretion diagnostic the H$\alpha$ $10\%$ width. By comparing this measurement with veiling in their stellar spectra, they found that a H$\alpha$ $10\%$ width larger than $\!270$~km~s$^{-1}$ reliably indicates accretion. A less conservative accretion cutoff of 200~km~s$^{-1}$ was adopted by \citet{Jayawardhana:2003p1130} for their study of young very low mass objects, based on empirical results and physical reasoning; however, they cautioned that it should be used in combination with additional diagnostics whenever possible. In later studies, it was found that H$\alpha$ $10\%$ width not only appears to be a good qualitative indicator of accretion but also correlates with the mass accretion rates derived by other means: the $200$~km~s$^{-1}$ threshold corresponds to a mass accretion rate of $\sim\!10^{-11}$ M$_{\sun}$ yr$^{-1}$ \citep{Natta:2004p2787}.

For this study, we use the H$\alpha$ $10\%$ width as one accretion diagnostic, which we computed as follows. First, we estimated the continuum level at H$\alpha$ by linearly interpolating between flux measurements in the range of $500$~km~s$^{-1}$ to $1000$~km~s$^{-1}$ on either side of the line. Next, the maximum flux level of H$\alpha$ emission was measured with respect to this continuum level. Finally, the crossing points of the H$\alpha$ emission with the $10\%$ flux level are identified, and the width was measured. The results for our targets are listed in tables \ref{tbl-SummaryCha} \& \ref{tbl-SummaryTau}; we note that for some objects, the measurements were uncertain, e.g.\ because of absorption components or double-peaked profiles with one peak close to 10\% of the height of the main peak. Note, however, this is not critical: all these sources are clearly accretors.

To obtain mass accretion rates, we use the \ion{Ca}{2}-$\lambda$8662 emission fluxes (${\mathcal F}_\textrm{\scriptsize{Ca}\,\tiny{I$\!$I}}$) which have been shown to be a more robust quantitative indicator of accretion than H$\alpha$ 10\% width (Nguyen et al. 2008, submitted to ApJL). We derived the fluxes from the observed emission equivalent widths. To determine the widths, we integrated the emission above the continuum level. For emission profiles attenuated by a broad absorption feature, we used the median flux within 0.2\,\AA\ of the absorption minima as an approximate continuum level for integration, similar to what was done by \citet{Muzerolle:1998p4143}. To infer the emission fluxes from the equivalent widths, we must know the underlying photospheric continuum flux. We used the continuum flux predicted by the PHOENIX synthetic spectra for a specified $T_{\rm eff}$ and surface gravity. We inferred $T_{\rm eff}$ from our spectral types, and assumed a surface gravity of $\log g = 4.0$ (cgs units). For five targets shared by \citet{Mohanty:2005p4012}, our results were lower by 0.05 to 0.41 dex. We ignored veiling, which may lead to an underestimate of line fluxes.

\subsection{Projected Rotational Velocity}
\label{sec-Rotation}

The projected rotational velocity $v~\!\sin~\!i$ of each target was determined by fitting the target spectra against sets of artificially broadened template spectra derived from one of the observed slowly rotating standard stars. For each target, we initially selected the standard star closest in spectral type as a template. To broaden the templates, we convolved the original template spectra with the analytical rotational broadening function of \citet{Gray:2005p2845} assuming a limb darkening factor of 0.65.

Our routine to estimate $v~\!\sin~\!i$ of a target consists of four steps. First, we fitted the target spectra with template spectra broadened from 0 to 200~km~s$^{-1}$ in steps of 10~km~s$^{-1}$, and recorded the $v~\!\sin~\!i$ value of the best fit for each echelle order. Second, we refined our search to projected rotational velocities within 10~km~s$^{-1}$ of the first-pass results in steps of 1~km~s$^{-1}$, and revised our estimates accordingly. Third, we computed weighted averages over the echelle orders, after removing outliers using a standard Tukey filter, i.e. values lying 1.5 times the interquartile range below the first quartile and above the third quartile were discarded (cf. \citet{CasterHoaglin:2000p3248}; for a Gaussian distribution, this filter corresponds to removing data points beyond $2.7$ $\sigma$). Fourth, we calculated the weighted average across epochs and used it as a provisional $v~\!\sin~\!i$ estimate of the target.

To finalize our $v~\!\sin~\!i$ estimates, we checked the provisional results using different templates and found that the variation in estimates was typically an order of magnitude larger than the weighted standard error of individual estimates. Therefore, for each target, we calculated two additional $v~\!\sin~\!i$ estimates using the next two closest standard stars by spectral type, and adopted as $v~\!\sin~\!i$ the estimate from the best-fit template, and as uncertainty, the standard deviation of the estimates between different templates. The results are listed in Tables \ref{tbl-SummaryCha} \& \ref{tbl-SummaryTau}. We considered the potential influence of veiling on our $v~\!\sin~\!i$ estimates: strong mass accretors will have strong veiling which could affect the $v~\!\sin~\!i$ estimates. However, we found no correlation between accretion signatures and rotational velocities when comparing these values for individual stars over time.

\section{Results}
\label{sec-Results}

\subsection{Accretion and Disk Presence}
\label{sec-Accretion-Disk}

\begin{figure}
\begin{center}
\includegraphics[width=8cm]{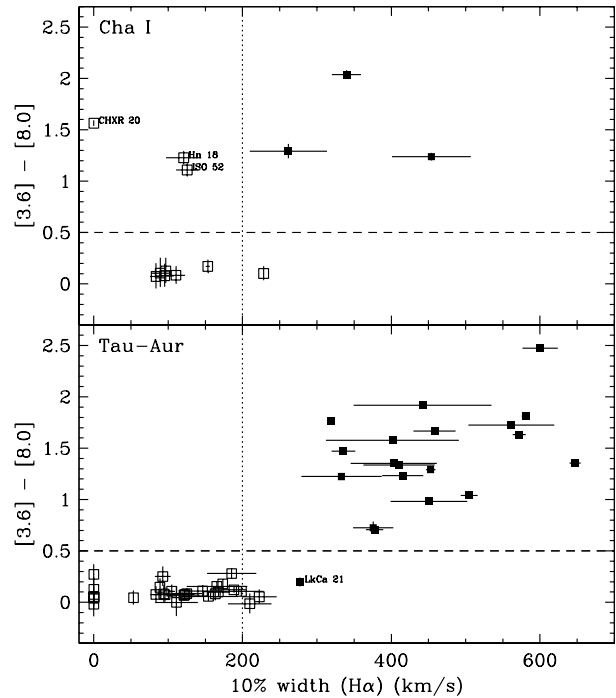}
\caption{\label{fig-M3680vsHa10w}The 8$\mu$m excess vs. the full width of H$\alpha$ at 10\% of the peak (H$\alpha$ 10\% width) for 13 Cha~I and 54 Tau-Aur members. Suspected accretors and non-accretors based on H$\alpha$ emission are denoted by solid and hollow symbols, respectively. The H$\alpha$ 10\% width error bars do not correspond to the measurement uncertainty, but to the scatter in our multi-epoch data. There is a clear separation of disk candidates above (and the non-disk candidates below) $[3.6] - [8.0] = 0.5$ illustrated by the dashed line, and a delineation between accretors and non-accretors at the cutoff of 200~km~s$^{-1}$ adopted originally by \citet{Jayawardhana:2003p1130}. Note that some non-accretors appear above the cutoff because of the additional broadening due to rapid rotation, see Fig.~\ref{fig-BoxVsiniHMvsLM}.}
\end{center}
\end{figure}

To examine the correlation between disk presence and accretion, we show in Fig.~\ref{fig-M3680vsHa10w} the $8\mu$m excess against H$\alpha$ $10\%$ widths for those targets for which both measurements are available. (The error bars on the H$\alpha$ $10\%$ width refer to the standard deviation of the estimates over epochs.) Out of the 67 objects in this subsample, 22 show evidence of both accretion and disk presence (cf. upper right regions in Fig.~\ref{fig-M3680vsHa10w}), implying that gas from the inner disk is still being channeled onto the star, and 41 objects have neither infrared excess nor signs of accretion. Thus, in nearly all cases $(63/67)$, the accretion signature is well correlated with disk presence.

Of the four exceptions, the three non-accretors with infrared excess, all in Cha~I, are CHXR~20, Hn~18, and ISO~52. For these objects, accretion rates may have dropped below measurable levels in H$\alpha$ 10\% width even though the disks persist. The \ion{Ca}{2}-$\lambda$8662 flux of Hn~18 is detectable and indicates a negligible accretion rate of $4.4 \times 10^{-11}\,M_{\sun}$\,yr$^{-1}$. Also, accretion may be variable on short timescales (Nguyen et al. 2008, submitted to ApJL). For CHXR~20, \ion{Ca}{2}-$\lambda$8662 emission was undetected at one epoch, and is present at two other epochs with a suggested small accretion rate of $2.6 \times 10^{-10}\,M_{\sun}$\,yr$^{-1}$. The only accretor without infrared excess, LkCa~21, was observed on a single epoch with a H$\alpha$ 10\% width of $277$\,km~s$^{-1}$; this value includes a contribution from rotational broadening of $46$\,km~s$^{-1}$. The net 10\% width is below the threshold for accretors originally set out by \citep{White:2003p1049}. In addition, \ion{Ca}{2}-$\lambda$8662 emission was not observed in LkCa~21 implying that it likely is not an accretor.

\subsection{Stellar Mass and Rotational Velocity}
\label{sec-Mass-Rotation}

\begin{figure}
\begin{center}
\includegraphics[width=8cm]{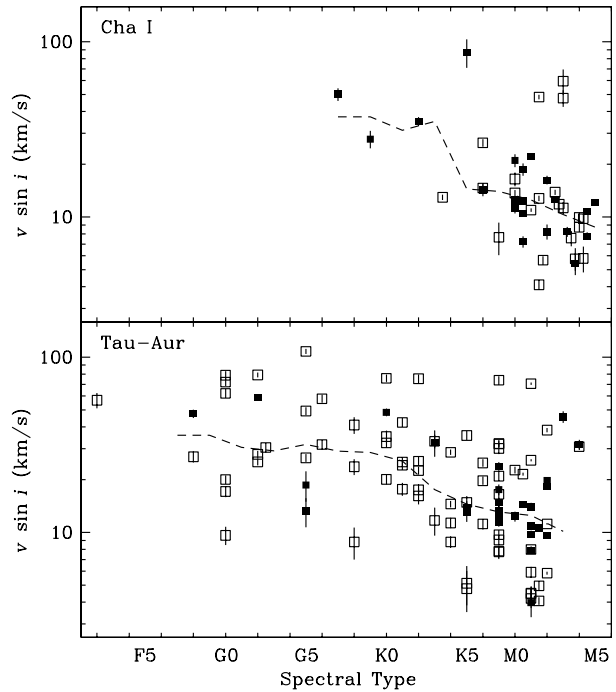}
\caption{\label{fig-Vsinivs}Projected rotational velocities $v~\!\sin~\!i$ as a function of spectral type. Suspected accretors and non-accretors based on H$\alpha$ emission are denoted by solid and hollow symbols, respectively. The $v~\!\sin~\!i$ errors represent the combined uncertainty between results using different template spectra, and over different epochs. The dashed line represents the median $v~\!\sin~\!i$ for bins covering on either side two spectral subtypes. The overall appearance of this plot is comparable to $v~\!\sin~\!i$ distribution in other young clusters: projected rotational velocity tends to increase with stellar mass.}
\end{center}
\end{figure}

Rotational velocity is known to vary as a function of stellar mass in young stars, likely because the efficiency of angular momentum removal depends on magnetic activity, which in turn depends on stellar mass \citep[cf.][]{Scholz:2007p1988}. To probe the rotation-mass dependence, we show the projected rotational velocity as a function of spectral type in Fig.~\ref{fig-Vsinivs}. Here late-K spectral type corresponds to $\sim\!1\,M_\sun$ \citep{Baraffe:1998p944}. The results are similar to what was found previously \citep[e.g.][]{Scholz:2007p1988,Rebull:2002p3923}. Higher mass stars tend to have faster projected rotational velocity overall than their lower mass counterparts.

\begin{figure}
\begin{center}
\includegraphics[width=8cm]{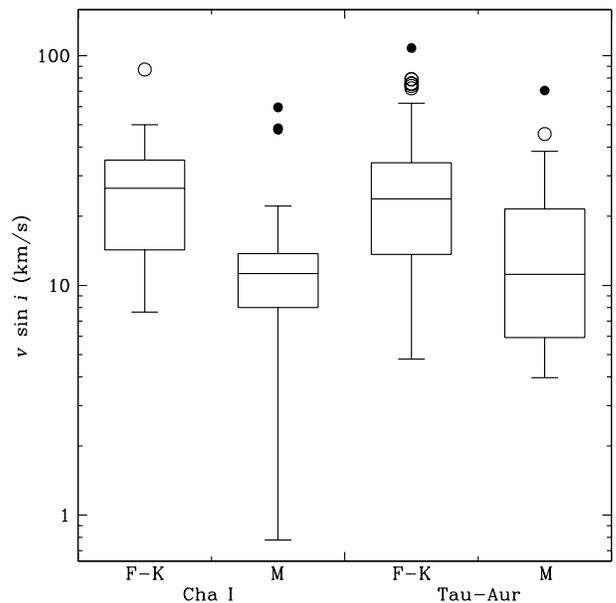}
\caption{\label{fig-BoxVsiniHMvsLM}Boxplots of $v~\!\sin~\!i$ for Cha~I and Tau-Aur grouped into two mass bins. Clearly, for both regions, the rotational velocities of high mass stars is faster than their lower mass counterparts by a factor of $2$--$2.5$. The central rectangles span the first quartile to the third quartiles with the segment inside indicating the median values, and ``whiskers'' above and below the box show the locations of the minima and maxima after applying a Tukey filter; statistical outliers and suspected outliers are shown as filled dots and hollow dots, respectively.}
\end{center}
\end{figure}

To examine this rotation-mass trend further, we divided our targets into two mass bins consisting of F--K type stars, and M type stars. In Fig.~\ref{fig-BoxVsiniHMvsLM}, we show boxplots of $v~\!\sin~\!i$ for the two mass bins: the horizontal lines inside the rectangles indicate the median values. Clearly, in both Cha~I and Tau-Aur, the median $v~\!\sin~\!i$ for the higher mass bins, 26~km~s$^{-1}$ and 24~km~s$^{-1}$, are significantly faster than those of the lower mass bins, both at 11~km~s$^{-1}$.

To get a quantitative sense of the difference in rotational velocity between the high and low mass stars, we applied the Kolmogorov-Smirnov (K-S) test. This analysis shows there is a probability of only $\sim\!0.5\%$ for Cha~I, and $\sim\!0.1\%$ for Tau-Aur that the $v~\!\sin~\!i$ for the two mass bins were drawn from the same distribution.

\begin{figure}
\begin{center}
\includegraphics[width=8cm]{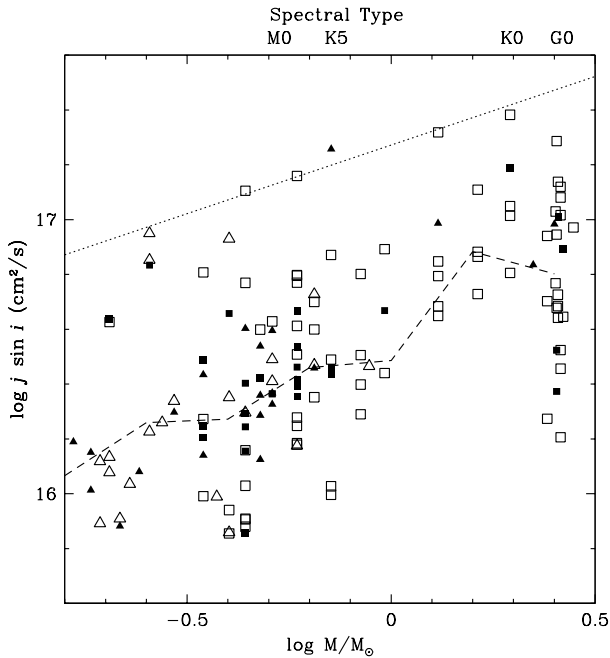}
\caption{\label{fig-JsinivsM}Specific angular momentum $j$ as a function of stellar mass computed assuming an age of $2$\,Myr from the models of \citet{DAntona:1997p3853}. Suspected accretors and non-accretors based on H$\alpha$ emission are denoted by solid and hollow symbols, respectively. Targets from Cha~I are represented by triangles, and those from Tau-Aur are drawn as squares. The dashed line represents the median $v~\!\sin~\!i$ for bins spanning $\log M/M_\odot \pm 0.1$. The dotted line is a linear fit by eye to the upper bound of the data and has a slope of $0.5$.}
\end{center}
\end{figure}

When interpreting this finding, one should take into account that the stars in our sample assuming an age of $\sim\!2$\,Myr span roughly a range of $1$--$4\,R_{\sun}$ in stellar radii. To gauge the contribution of stellar radius on $v~\!\sin~\!i$, we evaluated the specific angular momentum in our sample as follows. First, we converted spectral type to effective temperature by looking up and interpolating values from \citet{Sherry:2004p5936}. Second, we used the effective temperature to obtain estimates of mass $M$, radius $R$, and moment of inertia $I$ from the models of \citet{DAntona:1997p3853}. Third, we combined these values with our $v~\!\sin~\!i$ estimates to compute the projected specific angular momentum using the relation $j~\!\sin~\!i = (v~\!\sin~\!i)I/MR$. In Fig.~\ref{fig-JsinivsM}, we show $j~\!\sin~\!i$ as a function of stellar mass. From the best linear fit to the upper envelope of the datapoints, we find by eye that $j \propto M^{0.5}$. Indeed, there is an increase in specific angular momentum with increasing stellar mass.

\subsection{Accretion and Rotational Velocity}
\label{sec-Accretion-Rotation}

\begin{figure}
\begin{center}
\includegraphics[width=8cm]{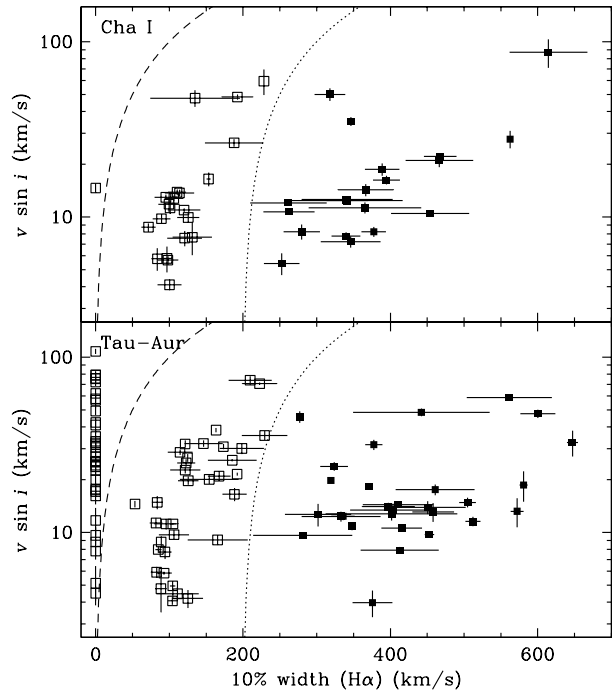}
\caption{\label{fig-VsinivsHa10w}Projected rotational velocity $v~\!\sin~\!i$ vs. H$\alpha$ 10\% width for a sample of T Tauri stars in the Chamaeleon~I and Taurus-Auriga star forming regions. Suspected accretors and non-accretors based on H$\alpha$ emission are denoted by solid and hollow symbols, respectively. The H$\alpha$ 10\% width error bars do not correspond to the measurement uncertainty, but to the scatter in our multi-epoch data. The $v~\!\sin~\!i$ errors represent the combined uncertainty between results using different template spectra, and over different epochs. The intrinsic contribution of rotation to line width is shown by the dashed line. The adopted boundary between accretors and non-accretors is shown by the dotted line.}
\end{center}
\end{figure}

\begin{figure}
\begin{center}
\includegraphics[width=8cm]{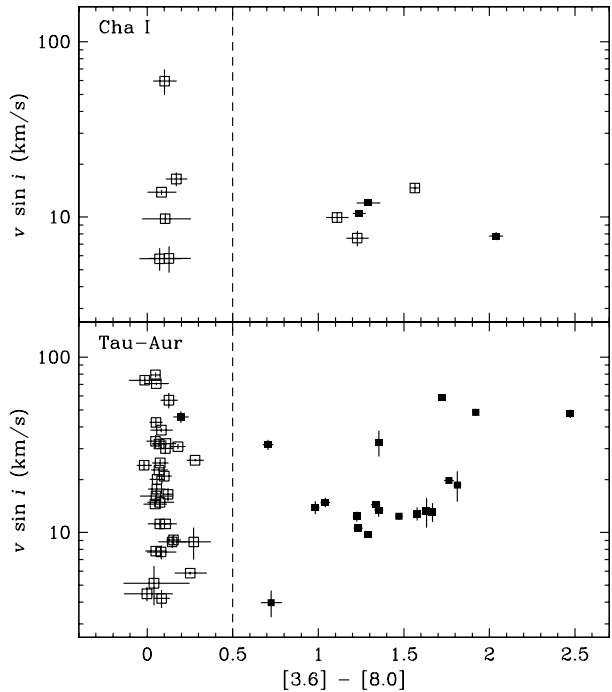}
\caption{\label{fig-VsinivsM3680}Projected rotational velocity $v~\!\sin~\!i$ as a function of 8$\mu$m excess. Suspected accretors and non-accretors based on H$\alpha$ emission are denoted by solid and hollow symbols, respectively. The $v~\!\sin~\!i$ errors represent the combined uncertainty between results using different template spectra, and over different epochs. The separation of disk and non-disk candidates is illustrated by the dashed line. The $v~\!\sin~\!i$ distributions for stars with and without inner disks are not statistically distinct.}
\end{center}
\end{figure}

To check for a connection between accretion and rotation, in Fig.~\ref{fig-VsinivsHa10w} \& \ref{fig-VsinivsM3680}, we show $v~\!\sin~\!i$ as a function of H$\alpha$ 10\% width and of $8\,\mu$m excess for our targets. In addition, the figures show both the intrinsic contribution of rotation to the line widths, and the separation between accretors and non-accretors.

\begin{figure}
\begin{center}
\includegraphics[width=8cm]{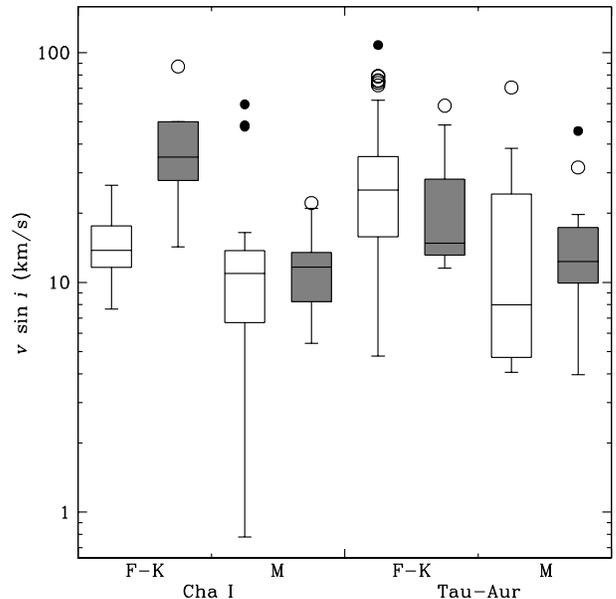}
\caption{\label{fig-BoxVsiniAccvsNonAcc}Boxplots of $v~\!\sin~\!i$ for accretors vs. non-accretors grouped by region and spectral type. The distributions in black and gray represent accretors and non-accretors, respectively. The central rectangles span the first quartile to the third quartiles with the segment inside indicating the median values, and ``whiskers'' above and below the box show the locations of the minima and maxima after applying a Tukey filter; statistical outliers and suspected outliers are shown as filled dots and hollow dots, respectively. The $v~\!\sin~\!i$ distributions between accretors and non-accretors are statistically similar. Note, the comparison for Cha~I high-mass stars involves only nine objects, and may appear deceivingly distinct by-eye.}
\end{center}
\end{figure}

We compared the distribution of $v~\!\sin~\!i$ for accretors and non-accretors using a number of K-S tests. To account for the rotation-mass dependence (see \S \ref{sec-Mass-Rotation}), we carried out these tests for the two mass bins (F--K type and M type) separately. The probability that the $v~\!\sin~\!i$ of accretors and non-accretors were drawn from the same distribution in Cha~I is $6\%$ for the high-mass targets, and $50\%$ for the low-mass ones. The probabilities for high- and low-mass targets in Tau-Aur are $8\%$ and $10\%$, respectively. For the entire sample, the corresponding probabilities for high- and low-mass targets is $30\%$ and $7\%$, respectively. Thus, any connection between accretion and projected rotational velocity is at best marginally significant. The $v~\sin~i$ distributions are shown in Fig.~\ref{fig-BoxVsiniAccvsNonAcc}.

Since the presence of dusty disks is strongly correlated with accretion in our targets, it is not surprising there is, for high and low mass stars in Tau-Aur respectively, a $9\%$ and $13\%$ probability that the $v~\!\sin~\!i$ for stars with and without disks were drawn from the same distribution. It would appear that the presence of ongoing accretion or a disk has no significant effect on the rotation in our sample. This is contrary to the standard disk braking scenario, as outlined in \S\ref{sec-Introduction}.

One particular reason for the negative test results is the presence of a significant number of rapidly rotating accretors, as seen in Fig.~\ref{fig-VsinivsHa10w}. Based on Spitzer data, \citet{Rebull:2006p190} find that the fraction of stars with disks is very low for rotation periods $P<1.8$\,d (see their Fig.~3). For a radius of 1$\,R_{\sun}$ and an average $\sin i$, this period corresponds to a projected rotational velocity of 22~km~s$^{-1}$. This value scales linearly with stellar radius. In our sample, we see 5--10 objects rotating faster than this threshold, where the exact number depends on the inclinations and stellar radii. This type of objects is not expected in the evolutionary sequence for the standard disk braking scenario described in \S\ref{sec-Introduction}.

Previous studies drew conclusions about the disk-braking scenario based on rotation periods from photometric data, while we use projected rotational velocity. To check whether this makes a difference, we ran Monte Carlo simulations based on published data from previous photometric studies, e.g.\ \citet{Stassun:1999p1017}, \citet{Herbst:2002p993} surveys in the ONC. In the simulations, rotation periods were converted to $v~\!\sin~\!i$ by selecting random viewing angles and using uniformly distributed stellar radii of $1$--$4 R_{\sun}$. In the case of \citet{Herbst:2002p993}, where there was previous indication of disk-braking, we found probabilities of $<\!1\%$ that diskless stars have the same distribution of $v~\!\sin~\!i$ as disk harboring stars, hence we recovered  their evidence for disk braking. Furthermore, for data from \citet{Stassun:1999p1017}, where disk-braking was not observed, we found that the simulated $v~\!\sin~\!i$ distributions for diskless and disk-harboring stars were similar, with probabilities consistently $>\!10\%$. We conclude that our results are insensitive to our use of projected rotational velocity to probe disk-braking scenarios.

\section{Summary and Discussion}
\label{sec-Summary}

We present a comprehensive study of projected rotational velocities and H$\alpha$ 10\% widths for young stars in Taurus-Auriga and Chamaeleon~I. Our three main results are:

1. {\it Indicators for accretion and inner disks agree for $>\!94$\% of our total sample.} For nearly all objects, the dissipation of the dusty inner disk and the drop in accretion rate below measurable levels occur simultaneously. Consequently, the lifetimes of inner disks are similar to the timescales of gas accretion. Based on our large sample, systems with inner disk clearings and ongoing gas accretion are rare (1/67); the same holds for systems with non-accreting inner disks (3/67). This consistency shows that timescales for inner disk clearing and accretion decline are much shorter than typical disk lifetimes ($\sim\!10^5$ years instead of several $10^6$ years).

2. {\it F--K stars have on average 2--2.5 times larger rotational velocities than M stars.} Although rotational velocity is proportional to stellar radius, from the models of \citet{DAntona:1997p3853} at $2$\,Myr, the typical radius of our F--K stars is less than 1.5 times that of the M stars in our sample. Moreover, the specific angular momentum is proportional to $M^{0.5}$. This mass dependence complements findings in Orion of \citet{Wolff:2004p4307} where the upper envelope of the observed values of angular momentum per unit mass varies as $M^{0.25}$ for stars on convective tracks ($\sim\!1$\,Myr) with a break in the power law with a sharp decline in $j$ with decreasing mass for stars with $M\!<\!2\,M_{\sun}$ for slightly older stars on radiative tracks (see their Fig.~3). They posit that these broad trends can be accounted for by simple models where stars lose angular momentum before they are deposited on the birth line, plausibly through star-disk interaction, and for stars with $M\!<\!2\,M_{\sun}$, the amount of braking increases with time spent evolving down their convective tracks. Our analysis of $\sim\!2$\,Myr old T Tauri stars in Cha~I and Tau-Aur showed an angular momentum mass trend in-between that of the two age groups studied in Orion for the same mass regime. This intermediate result could hint at only the beginning stages of disk braking, a significant age spread, or both.

3. {\it The presence of accretion or an inner disk does not significantly affect the distribution of rotational velocities in Taurus-Auriga and Chamaeleon~I.} This finding adds to the ongoing debate on disk braking in young stars, as it is in stark contrast to recent studies in the ONC and NGC~2264 \citep[e.g.][]{Rebull:2006p190,Cieza:2007p3508}, where a clear increase in disk fraction is found with increasing rotation period.

Part of the explanation may be that the stars have had insufficient time to brake. \citet{Hartmann:2002p2704} estimate a disk-magnetosphere braking timescale $\tau_{\rm DB}\,\gtrsim\,4.5\,\times\,10^6$\,yr\,$M_{0.5}\,\dot{M}^{-1}_{-8} f$, where $M_{0.5}$ is the stellar mass in units of $0.5\,M_{\sun}$, $\dot{M}_{-8}$ is the mass accretion rate in units of the typical value of $10^{-8}\,M_{\sun}$\,yr$^{-1}\,(M / 0.5 M_{\sun})$, and $f$ is the stellar rotation as a fraction of breakup velocity, then $\tau_{\rm DB}$ for our sample (typically $\sim\!5$\,Myr) is somewhat larger than the estimated age of the stars ($\sim\!2$\,Myr). Adopting the accretion and rotation results of \citet{Rebull:2000p4414} and \citet{Clarke:2000p4320}, we find a shorter typical $\tau_{\rm DB}$ of $\sim\!1$\,Myr for the ONC where disk braking is observed. Therefore, disk-locking in Cha~I and Tau-Aur may be ineffective overall.

In addition, there is evidence for spin-down not involving accretion from the inner disk. The specific angular momentum values for $\sim\!10$\,Myr solar-like mass stars in Orion, where strong braking is observed \citep{Wolff:2004p4307}, are typically lower than those found in our sample of similar mass non-accretors (where disk-locking should have expired) by a factor of $\sim\!5$. In contrast, the $j$ values for $\sim\!1$\,Myr solar-like mass stars in Orion are similar to those of our non-accretors. These measurements could imply another braking mechanism is at work after disks have dissipated. Of course, this indication for another braking mechanism relies on accurate age estimates.

Another possible explanation for the lack of strong disk braking in our sample is age spread, which might be larger in Taurus-Auriga and Chamaeleon~I than in the cluster cores of ONC and NGC~2264. Covering objects in varying stages of their rotational history could dilute a disk braking signature. In addition, the contrast between our results and those of previous studies that find strong evidence of disk braking may stem from different initial conditions at evolutionary stages before the stars became optically observable, e.g. at the birth line. The initial rotational velocity distributions for dense star forming regions like ONC and NGC~2264 could be very different from Tau-Aur and Cha~I where star formation has occurred in an environment with much lower stellar density. Future investigation of a similar kind should aim to take into account a more complete understanding of stellar properties when looking at correlations between rotation and disk/accretion signatures. In combination with previous results in other young clusters, our data will serve as an empirical basis for future studies on the timescales and efficiency of disk braking mechanisms.

\acknowledgments
We thank the anonymous referee for helpful comments that improved the clarity of the paper. We would like to thank David Lafreni\`{e}re, and Nairn Baliber for useful suggestions relating to the work presented in this paper. This paper includes data gathered with the 6.5 meter Magellan Telescopes located at Las Campanas Observatory, Chile. We would also like to thank the Magellan staff for their tireless effort and patience on accommodating our aggressive observing program. This work was supported in part by NSERC grants to RJ and MHvK, and an Early Researcher Award from the province of Ontario to RJ.

\bibliographystyle{apj}

\clearpage

\LongTables
\begin{landscape}
\begin{deluxetable}{llccccccccc}
\tablecaption{\label{tbl-SummaryCha}Summary of Results: Cha~I}
\tablewidth{0pt}
\tabletypesize{\scriptsize}
\tablehead{
& & \colhead{$v~\!\sin~\!i$\tablenotemark{a}} & \colhead{$10\%$ width\tablenotemark{b}} & \colhead{H$\alpha$ EW} & \colhead{\ion{Ca}{2} EW} & \colhead{S/N\tablenotemark{c}} & \colhead{$3.6\mu$m} & \colhead{$4.5\mu$m} & \colhead{$5.8\mu$m} & \colhead{$8.0\mu$m}\\
\colhead{Object} & \colhead{SpT} & \colhead {(km~s$^{-1}$)} & \colhead{(km~s$^{-1}$)} & \colhead{(\AA)} & \colhead{(\AA)} & \colhead{at H$\alpha$} & \colhead{(mJy)} & \colhead{(mJy)} & \colhead{(mJy)} & \colhead{(mJy)}
}
\startdata
\input{tab1.tex}
\enddata
\tablenotetext{a}{The H$\alpha$ 10\% width uncertainty does not correspond to the measurement uncertainty, but to the scatter in our multi-epoch data.}
\tablenotetext{b}{The $v~\!\sin~\!i$ uncertainty represents the combined measurement scatter between results using different template spectra, and over different epochs.}
\tablenotetext{c}{The S/N is based on the continuum on either side of H$\alpha$ used for the 10\% width calculations, and the uncertainty represents the standard error of the estimate.}
\end{deluxetable}

\clearpage
\end{landscape}

\LongTables
\begin{landscape}
\begin{deluxetable}{llccccccccc}
\tablecaption{\label{tbl-SummaryTau}Summary of Results: Tau-Aur}
\tablewidth{0pt}
\tabletypesize{\scriptsize}
\tablehead{
& & \colhead{$v~\!\sin~\!i$\tablenotemark{a}} & \colhead{$10\%$ width\tablenotemark{b}} & \colhead{H$\alpha$ EW} & \colhead{\ion{Ca}{2} EW} & \colhead{S/N\tablenotemark{c}} & \colhead{$3.6\mu$m} & \colhead{$4.5\mu$m} & \colhead{$5.8\mu$m} & \colhead{$8.0\mu$m}\\
\colhead{Object} & \colhead{SpT} & \colhead {(km~s$^{-1}$)} & \colhead{(km~s$^{-1}$)} & \colhead{(\AA)} & \colhead{(\AA)} & \colhead{at H$\alpha$} & \colhead{(mJy)} & \colhead{(mJy)} & \colhead{(mJy)} & \colhead{(mJy)}
}
\startdata
\input{tab2.tex}
\enddata
\tablenotetext{a}{The H$\alpha$ 10\% width uncertainty does not correspond to the measurement uncertainty, but to the scatter in our multi-epoch data.}
\tablenotetext{b}{The $v~\!\sin~\!i$ uncertainty represents the combined measurement scatter between results using different template spectra, and over different epochs.}
\tablenotetext{c}{The S/N is based on the continuum on either side of H$\alpha$ used for the 10\% width calculations, and the uncertainty represents the standard error of the estimate.}
\tablenotetext{d}{Spectral type determined by this work.}
\end{deluxetable}

\clearpage
\end{landscape}

\end{document}

%% file: tab1.tex
T4 & M0.5 & 12.4 $\pm$ 0.4 & 341 $\pm$ 28 & -15 $\pm$ 2 & -0.34 $\pm$ 0.12 & 14.7 $\pm$ 1.2 & \nodata & \nodata & \nodata & \nodata\\
T7 & K8 & 11.3 $\pm$ 0.8 & 365 $\pm$ 76 & -30 $\pm$ 9 & -1.6 $\pm$ 1.1 & 13.0 $\pm$ 1.0 & \nodata & \nodata & \nodata & \nodata\\
T8 & K2 & 35 $\pm$ 2 & 347 & -18 & -0.1 & 40 $\pm$ 5 & \nodata & \nodata & \nodata & \nodata\\
T10 & M3.75 & 5.4 $\pm$ 0.8 & 252 $\pm$ 24 & -90 $\pm$ 27 & -0.2 $\pm$ 0.02 & 5.7 $\pm$ 0.8 & \nodata & \nodata & \nodata & \nodata\\
T11 & K6 & 14.3 $\pm$ 1.1 & 367 $\pm$ 38 & -41 $\pm$ 4 & -0.318 $\pm$ 0.012 & 32 $\pm$ 2 & \nodata & \nodata & \nodata & \nodata\\
T12 & M4.5 & 10.7 $\pm$ 0.2 & 262 $\pm$ 35 & -38 $\pm$ 7 & \nodata & 4.6 $\pm$ 1.0 & \nodata & \nodata & \nodata & \nodata\\
ISO 52 & M4 & 9.9 $\pm$ 0.6 & 126 $\pm$ 15 & -5.4 $\pm$ 1.0 & \nodata & 4.9 $\pm$ 1.0 & 29.0 $\pm$ 1.0 & 24.0 $\pm$ 1.0 & 17 $\pm$ 2 & 18.4 $\pm$ 0.9\\
CHXR 14N & K8 & 13.7 $\pm$ 0.6 & 114 $\pm$ 20 & -2.0 $\pm$ 0.9 & -0.29 $\pm$ 0.02 & 14.3 $\pm$ 1.1 & \nodata & \nodata & \nodata & \nodata\\
CHXR 14S & M1.75 & 5.7 $\pm$ 0.3 & 98 $\pm$ 14 & -3.24 $\pm$ 0.19 & -0.39 $\pm$ 0.04 & 10.1 $\pm$ 1.0 & \nodata & \nodata & \nodata & \nodata\\
T16 & M3 & 11.3 $\pm$ 0.9 & 101 & -9 & -0.28 $\pm$ 0.04 & 3.5 $\pm$ 1.0 & \nodata & 26.0 $\pm$ 1.0 & \nodata & 21.0 $\pm$ 1.0\\
T20 & M1.5 & 48.3 $\pm$ 1.4 & 192 $\pm$ 21 & -3.8 $\pm$ 0.4 & \nodata & 14.6 $\pm$ 1.3 & \nodata & \nodata & \nodata & \nodata\\
Hn 5 & M4.5 & 7.8 $\pm$ 0.4 & 340 $\pm$ 20 & -56 $\pm$ 16 & -2.3 $\pm$ 0.9 & 5.6 $\pm$ 1.0 & 63 $\pm$ 2 & 66 $\pm$ 2 & 71 $\pm$ 2 & 94 $\pm$ 2\\
T22 & M3 & 60 $\pm$ 10 & 228 & -3 & \nodata & 5.9 $\pm$ 0.9 & 65 $\pm$ 2 & 42 $\pm$ 2 & 32 $\pm$ 2 & 16.3 $\pm$ 0.9\\
CHXR 20 & K6 & 14.6 $\pm$ 0.9 & Absorp. & 0.6 & -0.19 $\pm$ 0.02 & 12.7 $\pm$ 1.7 & 113 $\pm$ 2 & 86 $\pm$ 2 & 81 $\pm$ 3 & 109 $\pm$ 2\\
CHXR 74 & M4.25 & 5.8 $\pm$ 1.0 & 97 $\pm$ 7 & -5.5 $\pm$ 0.8 & -0.23 $\pm$ 0.11 & 4.3 $\pm$ 0.9 & 30 $\pm$ 2 & 20.0 $\pm$ 1.0 & 12 $\pm$ 2 & 7.7 $\pm$ 0.7\\
CHXR 21 & M3 & 48 $\pm$ 5 & 135 $\pm$ 61 & -4.0 $\pm$ 1.9 & \nodata & 4.6 $\pm$ 0.9 & \nodata & \nodata & \nodata & \nodata\\
T24 & M0.5 & 10.5 $\pm$ 0.4 & 454 $\pm$ 53 & -18 $\pm$ 7 & -0.26 $\pm$ 0.04 & 10.9 $\pm$ 1.2 & 98 $\pm$ 2 & 79 $\pm$ 2 & 70 $\pm$ 2 & 70 $\pm$ 2\\
T25 & M2.5 & 12.6 $\pm$ 0.3 & 341 $\pm$ 62 & -14 $\pm$ 3 & -0.13 $\pm$ 0.06 & 8.0 $\pm$ 1.0 & \nodata & \nodata & \nodata & \nodata\\
CHXR 76 & M4.25 & 9.8 $\pm$ 0.6 & 89 $\pm$ 12 & -5.8 $\pm$ 1.8 & -0.13 $\pm$ 0.07 & 3.2 $\pm$ 1.0 & 16.7 $\pm$ 0.9 & 11.6 $\pm$ 0.8 & 4 $\pm$ 2 & 4.2 $\pm$ 0.5\\
T33A & K3.5 & 12.9 $\pm$ 0.5 & 95 $\pm$ 15 & -0.6 $\pm$ 0.2 & -0.22 $\pm$ 0.04 & 15.9 $\pm$ 1.4 & \nodata & \nodata & \nodata & \nodata\\
T33B & G7 & 50 $\pm$ 4 & 318 $\pm$ 21 & -61 $\pm$ 12 & -2.3 $\pm$ 1.5 & 8.1 $\pm$ 1.1 & \nodata & \nodata & \nodata & \nodata\\
T34 & M3.75 & 5.8 $\pm$ 0.8 & 84 $\pm$ 8 & -2.5 $\pm$ 1.1 & \nodata & 5.6 $\pm$ 0.8 & 34 $\pm$ 2 & 23.0 $\pm$ 1.0 & 16 $\pm$ 2 & 8.3 $\pm$ 0.8\\
T35 & K8 & 21.0 $\pm$ 1.8 & 466 $\pm$ 46 & -91 $\pm$ 49 & -0.35 $\pm$ 0.09 & 7.7 $\pm$ 0.9 & \nodata & 100 $\pm$ 2 & \nodata & 45 $\pm$ 2\\
CHXR 33 & M0 & 16.5 $\pm$ 1.5 & 153 $\pm$ 3 & -2.9 $\pm$ 0.2 & -0.18 $\pm$ 0.03 & 15.0 $\pm$ 1.5 & 67 $\pm$ 2 & 45 $\pm$ 2 & 36 $\pm$ 2 & 17.9 $\pm$ 0.9\\
T38 & M0.5 & 18.7 $\pm$ 1.5 & 389 $\pm$ 23 & -107 $\pm$ 37 & -0.7 $\pm$ 0.7 & 5.6 $\pm$ 1.0 & \nodata & \nodata & \nodata & \nodata\\
T39A sw & K7 & 7.7 $\pm$ 1.6 & 131 $\pm$ 27 & -5.2 $\pm$ 1.4 & -0.41 $\pm$ 0.09 & 11.9 $\pm$ 1.0 & \nodata & \nodata & \nodata & \nodata\\
T39A w & M1.5 & 4.1 $\pm$ 0.3 & 100 $\pm$ 16 & -2.6 $\pm$ 0.6 & -0.26 $\pm$ 0.06 & 9.8 $\pm$ 1.0 & \nodata & \nodata & \nodata & \nodata\\
T39B e & M1.5 & 12.8 $\pm$ 0.4 & 106 $\pm$ 13 & -5.3 $\pm$ 1.2 & -0.25 $\pm$ 0.03 & 7.2 $\pm$ 0.8 & \nodata & \nodata & \nodata & \nodata\\
Hn 10E & M3.25 & 8.2 $\pm$ 0.5 & 377 $\pm$ 17 & -62 $\pm$ 3 & -1.6 $\pm$ 0.8 & 4.3 $\pm$ 1.0 & \nodata & \nodata & \nodata & \nodata\\
T44 & K5 & 87 $\pm$ 16 & 614 $\pm$ 53 & -67 $\pm$ 13 & -27 $\pm$ 10 & 15.9 $\pm$ 1.5 & \nodata & \nodata & \nodata & \nodata\\
T45A & M0 & 12.4 $\pm$ 0.5 & 340 $\pm$ 77 & -2.7 $\pm$ 0.8 & -0.33 $\pm$ 0.03 & 16.3 $\pm$ 1.7 & \nodata & \nodata & \nodata & \nodata\\
T47 & M2 & 16.2 $\pm$ 0.9 & 395 $\pm$ 18 & -42 $\pm$ 7 & -2 $\pm$ 2 & 3.1 $\pm$ 1.0 & \nodata & \nodata & \nodata & \nodata\\
CHXR 48 & M2.5 & 13.8 $\pm$ 0.5 & 110 $\pm$ 12 & -4.3 $\pm$ 1.9 & -0.4 $\pm$ 0.3 & 9.2 $\pm$ 1.0 & 45 $\pm$ 2 & 29.0 $\pm$ 1.0 & 19 $\pm$ 2 & 11.1 $\pm$ 0.7\\
T49 & M2 & 8.2 $\pm$ 0.8 & 280 $\pm$ 25 & -87 $\pm$ 19 & -0.6 $\pm$ 0.2 & 6.3 $\pm$ 0.8 & \nodata & 120 $\pm$ 3 & \nodata & 117 $\pm$ 2\\
CHX 18N & K6 & 26.5 $\pm$ 1.3 & 188 $\pm$ 39 & -3.3 $\pm$ 0.7 & -0.26 $\pm$ 0.06 & 30 $\pm$ 2 & \nodata & 358 $\pm$ 4 & \nodata & 196 $\pm$ 3\\
T50 & M5 & 12.0 $\pm$ 0.4 & 262 $\pm$ 52 & -22 $\pm$ 3 & -0.1627 $\pm$ 0.0007 & 5.9 $\pm$ 0.9 & 52 $\pm$ 2 & 44 $\pm$ 2 & 35 $\pm$ 2 & 39 $\pm$ 2\\
T52 & G9 & 28 $\pm$ 3 & 562 & -48 & -8 & 35 $\pm$ 4 & \nodata & \nodata & \nodata & \nodata\\
T53 & M1 & 22.2 $\pm$ 0.6 & 468 $\pm$ 22 & -62 $\pm$ 20 & -3.5 $\pm$ 1.5 & 7.4 $\pm$ 0.9 & \nodata & \nodata & \nodata & \nodata\\
CHXR 54 & M1 & 10.9 $\pm$ 0.2 & 120 $\pm$ 22 & -1.3 $\pm$ 0.3 & -0.21 $\pm$ 0.03 & 17.1 $\pm$ 1.3 & \nodata & \nodata & \nodata & \nodata\\
Hn 17 & M4 & 8.7 $\pm$ 0.5 & 72 $\pm$ 10 & -2.6 $\pm$ 0.4 & \nodata & 4.9 $\pm$ 0.9 & \nodata & \nodata & \nodata & \nodata\\
CHXR 57 & M2.75 & 11.8 $\pm$ 1.2 & 100 $\pm$ 15 & -3.1 $\pm$ 1.1 & -0.29 $\pm$ 0.09 & 9.2 $\pm$ 1.0 & \nodata & \nodata & \nodata & \nodata\\
Hn 18 & M3.5 & 7.6 $\pm$ 0.8 & 121 $\pm$ 24 & -6.0 $\pm$ 1.7 & -0.101 $\pm$ 0.018 & 5.4 $\pm$ 0.7 & 27.0 $\pm$ 1.0 & 24.0 $\pm$ 1.0 & 20 $\pm$ 2 & 19.1 $\pm$ 0.9\\
CHXR 60 & M4.25 & 0.8 $\pm$ 0.7 & 95 $\pm$ 12 & -5.8 $\pm$ 1.1 & -0.1 & 4.5 $\pm$ 0.9 & 22.0 $\pm$ 1.0 & 17.1 $\pm$ 0.9 & 11 $\pm$ 2 & 5.4 $\pm$ 0.5\\
T56 & M0.5 & 7.2 $\pm$ 0.5 & 346 $\pm$ 41 & -49 $\pm$ 11 & -0.49 $\pm$ 0.17 & 15.2 $\pm$ 1.1 & \nodata & \nodata & \nodata & \nodata\\

%% file: tab2.tex
NTTS 034903+2431 & K5 & 36 $\pm$ 2 & 229 $\pm$ 31 & -1.6 $\pm$ 0.2 & \nodata & 24.9 $\pm$ 1.7 & \nodata & \nodata & \nodata & \nodata\\
NTTS 035120+3154SW & G0 & 62 $\pm$ 3 & Absorp. & 2.1 $\pm$ 0.3 & \nodata & 22.7 $\pm$ 1.5 & \nodata & \nodata & \nodata & \nodata\\
HD 285281 & K0 & 76 $\pm$ 3 & Absorp. & 0.15 $\pm$ 0.08 & \nodata & 48 $\pm$ 3 & \nodata & \nodata & \nodata & \nodata\\
NTTS 040047+2603E & M2 & 5.85 $\pm$ 0.11 & 93 $\pm$ 11 & -3.7 $\pm$ 0.9 & -0.18 $\pm$ 0.08 & 14.8 $\pm$ 0.9 & 22.3 $\pm$ 0.7 & 15.9 $\pm$ 0.7 & 11.9 $\pm$ 0.7 & 6.4 $\pm$ 0.5\\
RX J0405.1+2632 & K2 & 17.5 $\pm$ 1.3 & Absorp. & 0.7 $\pm$ 0.2 & -0.089 $\pm$ 0.004 & 29.7 $\pm$ 1.9 & \nodata & \nodata & \nodata & \nodata\\
RX J0405.3+2009 & K1 & 24.1 $\pm$ 1.4 & Absorp. & 0.64 $\pm$ 0.04 & -0.084 $\pm$ 0.011 & 41 $\pm$ 3 & 81.1 $\pm$ 1.2 & 48.9 $\pm$ 1.0 & 133.1 $\pm$ 1.0 & 18.2 $\pm$ 0.7\\
HD 284135 & G0 & 72 $\pm$ 4 & Absorp. & 2.03 $\pm$ 0.14 & \nodata & 58 $\pm$ 3 & \nodata & \nodata & \nodata & \nodata\\
HD 284149 & F8 & 27.0 $\pm$ 1.9 & Absorp. & 2.45 $\pm$ 0.10 & \nodata & 51 $\pm$ 3 & \nodata & \nodata & \nodata & \nodata\\
RX J0407.8+1750 & K4 & 28.7 $\pm$ 1.0 & 115 $\pm$ 17 & -0.59 $\pm$ 0.15 & -0.14 $\pm$ 0.02 & 24.7 $\pm$ 1.7 & \nodata & \nodata & \nodata & \nodata\\
RX J0408.2+1956 & K2 & 75 $\pm$ 4 & Absorp. & -0.1 & \nodata & 20.8 $\pm$ 1.4 & \nodata & \nodata & \nodata & \nodata\\
RX J0409.1+2901 & G8 & 24 $\pm$ 2 & Absorp. & 0.3 $\pm$ 0.07 & -0.12 $\pm$ 0.02 & 37 $\pm$ 2 & \nodata & \nodata & \nodata & \nodata\\
RX J0409.2+1716 & M1 & 70.5 $\pm$ 1.1 & 223 $\pm$ 23 & -3.9 $\pm$ 0.7 & \nodata & 14.9 $\pm$ 1.0 & 36.1 $\pm$ 0.8 & 22.7 $\pm$ 0.7 & 15.5 $\pm$ 0.8 & 8.7 $\pm$ 0.5\\
RX J0409.8+2446 & M1 & 5.9 $\pm$ 0.4 & 83 $\pm$ 8 & -1.9 $\pm$ 0.7 & -0.22 $\pm$ 0.05 & 19.6 $\pm$ 1.4 & \nodata & \nodata & \nodata & \nodata\\
RX J0412.8+1937 & K6 & 11.2 $\pm$ 0.8 & 96 $\pm$ 16 & -0.34 $\pm$ 0.09 & -0.19 $\pm$ 0.05 & 18.9 $\pm$ 1.2 & 35.5 $\pm$ 0.8 & 23.0 $\pm$ 0.7 & 14.9 $\pm$ 0.8 & 8.7 $\pm$ 0.5\\
HD 285579 & G0 & 9.6 $\pm$ 1.1 & Absorp. & 1.9 $\pm$ 0.5 & -0.066 $\pm$ 0.010 & 26.6 $\pm$ 1.5 & \nodata & \nodata & \nodata & \nodata\\
LkCa 1 & M4 & 30.9 $\pm$ 1.1 & 173 & -4 & \nodata & 24 $\pm$ 3 & 54.0 $\pm$ 1.0 & 31.5 $\pm$ 0.8 & 23.6 $\pm$ 0.8 & 14.6 $\pm$ 0.5\\
CW Tau & K3 & 33 $\pm$ 5 & 647 $\pm$ 7 & -87 $\pm$ 45 & -9 $\pm$ 3 & 21.0 $\pm$ 1.4 & 719 $\pm$ 5 & 694 $\pm$ 5 & 632 $\pm$ 5 & 572 $\pm$ 3\\
FP Tau & M4 & 32 $\pm$ 2 & 378 $\pm$ 12 & -12 $\pm$ 3 & \nodata & 14.1 $\pm$ 1.0 & 84.8 $\pm$ 1.4 & 63.8 $\pm$ 1.2 & 53.5 $\pm$ 1.2 & 37.1 $\pm$ 0.9\\
CX Tau & M2 & 19.8 $\pm$ 0.6 & 319 & -13 & -0.08 & 34 $\pm$ 5 & 56.6 $\pm$ 1.0 & 47.1 $\pm$ 1.0 & 48.2 $\pm$ 1.2 & 65.5 $\pm$ 1.2\\
RX J0415.3+2044 & K0 & 35 $\pm$ 3 & Absorp. & 1.16 $\pm$ 0.08 & \nodata & 37 $\pm$ 2 & \nodata & \nodata & \nodata & \nodata\\
RX J0415.8+3100 & G6 & 31.7 $\pm$ 1.9 & Absorp. & 1.6 $\pm$ 0.4 & \nodata & 20.2 $\pm$ 1.5 & \nodata & \nodata & \nodata & \nodata\\
LkCa 4 & K7 & 30 $\pm$ 2 & 198 $\pm$ 30 & -4.8 $\pm$ 0.9 & -0.23 $\pm$ 0.08 & 17.4 $\pm$ 1.3 & 72.3 $\pm$ 1.2 & 47.4 $\pm$ 1.0 & 32.4 $\pm$ 1.2 & 18.2 $\pm$ 0.7\\
CY Tau & M1.5 & 10.6 $\pm$ 0.4 & 415 $\pm$ 28 & -78 $\pm$ 13 & -1.0 $\pm$ 0.7 & 29 $\pm$ 2 & 89.5 $\pm$ 1.4 & 77.7 $\pm$ 1.2 & 67.4 $\pm$ 1.4 & 63.6 $\pm$ 1.2\\
LkCa 5 & M2 & 38.3 $\pm$ 1.1 & 163 & -4 & \nodata & 20 $\pm$ 3 & 38.0 $\pm$ 0.8 & 24.9 $\pm$ 0.7 & 16.5 $\pm$ 0.7 & 9.4 $\pm$ 0.5\\
NTTS 041529+1652 & K5 & 5.1 $\pm$ 1.3 & Absorp. & 0.3 $\pm$ 0.2 & -0.17 $\pm$ 0.04 & 14.8 $\pm$ 1.1 & 8.8 $\pm$ 0.5 & 5.8 $\pm$ 0.5 & 4.1 $\pm$ 0.7 & 2.1 $\pm$ 0.3\\
Hubble 4 & K7 & 16.5 $\pm$ 1.4 & 188 $\pm$ 16 & -3.2 $\pm$ 0.6 & -0.216 $\pm$ 0.017 & 20.2 $\pm$ 1.4 & 201 $\pm$ 3 & 135.5 $\pm$ 1.7 & 89.7 $\pm$ 1.5 & 51.3 $\pm$ 1.0\\
NTTS 041559+1716 & K7 & 74 $\pm$ 4 & 210 $\pm$ 29 & -1.5 $\pm$ 0.6 & \nodata & 20.8 $\pm$ 1.3 & 27.7 $\pm$ 0.8 & 18.0 $\pm$ 0.7 & 13.1 $\pm$ 0.7 & 6.2 $\pm$ 0.5\\
BP Tau & K5 & 13.1 $\pm$ 1.6 & 458 $\pm$ 28 & -109 $\pm$ 9 & -3.2 $\pm$ 0.6 & 24.6 $\pm$ 1.7 & 155.4 $\pm$ 1.7 & 135.5 $\pm$ 1.7 & 117.5 $\pm$ 1.5 & 164.7 $\pm$ 1.7\\
V819 Tau & K7 & 9.1 $\pm$ 0.6 & 166 $\pm$ 41 & -2.1 $\pm$ 1.5 & -0.22 $\pm$ 0.09 & 18.4 $\pm$ 1.3 & 73.1 $\pm$ 1.2 & 47.8 $\pm$ 1.0 & 35.0 $\pm$ 1.0 & 19.2 $\pm$ 0.7\\
DE Tau & M1 & 9.7 $\pm$ 0.3 & 453 $\pm$ 6 & -53 $\pm$ 9 & -5.2 $\pm$ 1.8 & 17.9 $\pm$ 1.3 & 208 $\pm$ 3 & 179.7 $\pm$ 1.7 & 149.4 $\pm$ 1.7 & 156.0 $\pm$ 1.7\\
RY Tau & F8 & 48 $\pm$ 3 & 600 $\pm$ 24 & -12 $\pm$ 4 & -0.8 $\pm$ 1.1 & 58 $\pm$ 3 & 1562 $\pm$ 7 & 1675 $\pm$ 7 & 2311 $\pm$ 7 & 3479 $\pm$ 9\\
HD 283572 & G2 & 79 $\pm$ 3 & Absorp. & 1.01 $\pm$ 0.14 & \nodata & 75 $\pm$ 4 & 253 $\pm$ 3 & 160.9 $\pm$ 1.7 & 108.7 $\pm$ 1.7 & 60.5 $\pm$ 1.2\\
LkCa 21 & M3 & 46 $\pm$ 3 & 277 & -5 & \nodata & 24 $\pm$ 3 & 68.4 $\pm$ 1.2 & 44.7 $\pm$ 1.0 & 31.8 $\pm$ 1.0 & 18.7 $\pm$ 0.7\\
HD 285751 & G5 & 26.6 $\pm$ 1.4 & 125 & -0.3 $\pm$ 0.18 & -0.119 $\pm$ 0.015 & 27.8 $\pm$ 1.7 & \nodata & \nodata & \nodata & \nodata\\
BD +26 718 & K0 & 32.4 $\pm$ 1.8 & Absorp. & 0.4 $\pm$ 0.2 & -0.06 $\pm$ 0.02 & 29 $\pm$ 2 & \nodata & \nodata & \nodata & \nodata\\
IP Tau & M0 & 12.3 $\pm$ 0.8 & 333 $\pm$ 54 & -15 $\pm$ 8 & -0.45 $\pm$ 0.07 & 19.2 $\pm$ 1.3 & 105.2 $\pm$ 1.4 & 90.1 $\pm$ 1.4 & 73.2 $\pm$ 1.4 & 74.2 $\pm$ 1.2\\
BD +17 724B & G5 & 49 $\pm$ 3 & Absorp. & 2.16 $\pm$ 0.08 & \nodata & 51 $\pm$ 3 & \nodata & \nodata & \nodata & \nodata\\
NTTS 042417+1744 & K1 & 17.6 $\pm$ 1.5 & Absorp. & 1.1 $\pm$ 0.3 & -0.12 $\pm$ 0.03 & 36 $\pm$ 2 & 64.2 $\pm$ 1.2 & 40.6 $\pm$ 0.8 & 26.0 $\pm$ 0.8 & 15.4 $\pm$ 0.7\\
DH Tau & M1 & 10.9 $\pm$ 0.6 & 348 & -59 & -2 & 29 $\pm$ 4 & \nodata & \nodata & \nodata & \nodata\\
IQ Tau & M0.5 & 14.4 $\pm$ 0.3 & 411 $\pm$ 48 & -25 $\pm$ 10 & -0.9 $\pm$ 0.8 & 19.0 $\pm$ 1.2 & 226 $\pm$ 3 & 213 $\pm$ 3 & 178 $\pm$ 3 & 177 $\pm$ 3\\
FX Tau a & M2\tablenotemark{d} & 9.61 $\pm$ 0.19 & 281 $\pm$ 67 & -6 $\pm$ 2 & -0.23 $\pm$ 0.13 & 12.5 $\pm$ 1.0 & \nodata & \nodata & \nodata & \nodata\\
FX Tau b & M1\tablenotemark{d} & 7.9 $\pm$ 0.3 & 413 $\pm$ 53 & -21 $\pm$ 12 & -0.3 $\pm$ 0.06 & 16.3 $\pm$ 1.2 & \nodata & \nodata & \nodata & \nodata\\
DK Tau A & K7 & 17.5 $\pm$ 1.2 & 461 $\pm$ 54 & -36 $\pm$ 27 & -2.2 $\pm$ 0.9 & 25.4 $\pm$ 1.8 & \nodata & \nodata & \nodata & \nodata\\
DK Tau B & M1\tablenotemark{d} & 14.0 $\pm$ 0.8 & 397 $\pm$ 35 & -46 $\pm$ 16 & -1.6 $\pm$ 1.1 & 10.5 $\pm$ 0.9 & \nodata & \nodata & \nodata & \nodata\\
RX J0430.8+2113 & G8 & 41 $\pm$ 4 & Absorp. & 0.5 $\pm$ 0.4 & \nodata & 45 $\pm$ 3 & \nodata & \nodata & \nodata & \nodata\\
HD 284496 & G0 & 20.0 $\pm$ 1.0 & Absorp. & 0.88 $\pm$ 0.11 & -0.08 $\pm$ 0.02 & 35 $\pm$ 2 & \nodata & \nodata & \nodata & \nodata\\
NTTS 042835+1700 & K5 & 14.8 $\pm$ 1.3 & 84 $\pm$ 7 & -0.36 $\pm$ 0.15 & -0.17 $\pm$ 0.03 & 22.9 $\pm$ 1.6 & 22.0 $\pm$ 0.7 & 13.9 $\pm$ 0.5 & 10.4 $\pm$ 0.7 & 5.4 $\pm$ 0.3\\
V710 Tau A & M0.5 & 21.5 $\pm$ 0.4 & 192 & -3 & \nodata & 13.5 $\pm$ 1.9 & \nodata & \nodata & \nodata & \nodata\\
V710 Tau B & M2 & 18.31 $\pm$ 0.19 & 371 & -37 & -0.4 & 16 $\pm$ 2 & \nodata & \nodata & \nodata & \nodata\\
L1551-51 & K7 & 32.1 $\pm$ 1.4 & 146 $\pm$ 26 & -1.0 $\pm$ 0.4 & -0.16 $\pm$ 0.02 & 21.6 $\pm$ 1.5 & 41.9 $\pm$ 0.8 & 26.8 $\pm$ 0.8 & 16.0 $\pm$ 0.7 & 10.6 $\pm$ 0.5\\
V827 Tau & K7 & 20.9 $\pm$ 1.3 & 168 $\pm$ 15 & -4.4 $\pm$ 1.3 & -0.26 $\pm$ 0.07 & 16.6 $\pm$ 1.2 & 74.8 $\pm$ 1.2 & 48.8 $\pm$ 1.0 & 33.3 $\pm$ 0.8 & 18.7 $\pm$ 0.7\\
GG Tau A a & K7 & 11.5 $\pm$ 0.7 & 512 $\pm$ 10 & -51 $\pm$ 4 & -2.5 $\pm$ 1.0 & 21.2 $\pm$ 1.6 & \nodata & \nodata & \nodata & \nodata\\
RX J0432.7+1853 & K1 & 25.2 $\pm$ 1.6 & Absorp. & 0.66 $\pm$ 0.16 & -0.09 $\pm$ 0.03 & 34 $\pm$ 2 & \nodata & \nodata & \nodata & \nodata\\
L1551-55 & K7 & 7.7 $\pm$ 0.7 & 94 $\pm$ 9 & -1.2 $\pm$ 0.4 & -0.32 $\pm$ 0.10 & 18.4 $\pm$ 1.4 & 28.9 $\pm$ 0.8 & 18.0 $\pm$ 0.7 & 12.6 $\pm$ 0.7 & 7.1 $\pm$ 0.5\\
RX J0432.8+1735 & M2 & 11.18 $\pm$ 0.11 & 105 $\pm$ 4 & -1.8 $\pm$ 0.3 & -0.28 $\pm$ 0.07 & 18.0 $\pm$ 1.2 & 37.8 $\pm$ 0.8 & 23.7 $\pm$ 0.7 & 16.6 $\pm$ 0.8 & 9.5 $\pm$ 0.5\\
V830 Tau & K7 & 32.0 $\pm$ 1.5 & 121 & -2 & -0.2 & 15 $\pm$ 2 & 59.1 $\pm$ 1.0 & 36.9 $\pm$ 0.8 & 24.8 $\pm$ 0.8 & 14.4 $\pm$ 0.5\\
GI Tau & K7 & 12.7 $\pm$ 1.9 & 302 $\pm$ 45 & -14 $\pm$ 5 & -0.62 $\pm$ 0.19 & 15.5 $\pm$ 1.3 & \nodata & \nodata & \nodata & \nodata\\
RX J0433.5+1916 & G6 & 58 $\pm$ 3 & Absorp. & 1.5 $\pm$ 0.3 & -0.04 & 17.8 $\pm$ 1.2 & \nodata & \nodata & \nodata & \nodata\\
DL Tau & G & 19 $\pm$ 4 & 581 $\pm$ 6 & -96 $\pm$ 9 & -41 $\pm$ 4 & 17.7 $\pm$ 1.3 & 233 $\pm$ 3 & 254 $\pm$ 3 & 246 $\pm$ 3 & 283 $\pm$ 3\\
DM Tau & M1 & 4.0 $\pm$ 0.7 & 376 $\pm$ 27 & -126 $\pm$ 37 & -0.26 $\pm$ 0.03 & 14.8 $\pm$ 1.2 & 24.1 $\pm$ 0.7 & 15.9 $\pm$ 0.7 & 11.2 $\pm$ 0.7 & 10.7 $\pm$ 0.5\\
CI Tau & G & 13 $\pm$ 2 & 572 $\pm$ 9 & -78 $\pm$ 7 & -23 $\pm$ 9 & 23.1 $\pm$ 1.7 & 225 $\pm$ 3 & 217 $\pm$ 5 & 188 $\pm$ 3 & 231 $\pm$ 3\\
HBC 407 & G8 & 8.8 $\pm$ 1.8 & Absorp. & 0.9 & -0.09 & 12 $\pm$ 2 & 15.4 $\pm$ 0.7 & 10.2 $\pm$ 0.5 & 6.1 $\pm$ 0.7 & 4.5 $\pm$ 0.3\\
AA Tau & K7 & 12.8 $\pm$ 1.1 & 402 $\pm$ 89 & -14 $\pm$ 8 & -0.34 $\pm$ 0.04 & 21.1 $\pm$ 1.5 & 172 $\pm$ 3 & 166.0 $\pm$ 1.7 & 152.8 $\pm$ 1.7 & 168.1 $\pm$ 1.7\\
HBC 412 A+B e & M1.5\tablenotemark{d} & 4.1 $\pm$ 0.2 & 104 $\pm$ 6 & -3.3 $\pm$ 1.2 & -0.157 $\pm$ 0.008 & 15.0 $\pm$ 1.5 & \nodata & \nodata & \nodata & \nodata\\
HBC 412 A+B w & M1.5\tablenotemark{d} & 4.9 $\pm$ 0.3 & 105 $\pm$ 4 & -4.0 $\pm$ 1.3 & -0.23 $\pm$ 0.03 & 14.4 $\pm$ 1.5 & \nodata & \nodata & \nodata & \nodata\\
DN Tau & M0 & 12.3 $\pm$ 0.6 & 336 $\pm$ 16 & -35 $\pm$ 6 & -0.42 $\pm$ 0.05 & 20.4 $\pm$ 1.3 & 135.1 $\pm$ 1.7 & 118.5 $\pm$ 1.7 & 108.7 $\pm$ 1.7 & 119.6 $\pm$ 1.7\\
HQ Tau & K0\tablenotemark{d} & 48 $\pm$ 2 & 442 $\pm$ 93 & -2.0 $\pm$ 0.6 & -0.03 $\pm$ 0.05 & 26.7 $\pm$ 1.8 & 378 $\pm$ 3 & 364 $\pm$ 3 & 367 $\pm$ 3 & 506 $\pm$ 3\\
RX J0435.9+2352 & M1 & 4.2 $\pm$ 0.5 & 125 $\pm$ 21 & -2.7 $\pm$ 1.5 & -0.21 $\pm$ 0.07 & 18.8 $\pm$ 1.4 & 52.7 $\pm$ 1.0 & 32.9 $\pm$ 0.8 & 21.9 $\pm$ 0.8 & 13.0 $\pm$ 0.5\\
LkCa 14 & M0 & 22.7 $\pm$ 1.0 & 122 $\pm$ 20 & -0.42 $\pm$ 0.15 & -0.17 $\pm$ 0.03 & 27.0 $\pm$ 1.6 & 52.7 $\pm$ 1.0 & 32.9 $\pm$ 0.8 & 21.7 $\pm$ 0.8 & 12.8 $\pm$ 0.5\\
HD 283759 & F2 & 57 $\pm$ 6 & Absorp. & 3.92 $\pm$ 0.10 & \nodata & 46 $\pm$ 3 & 62.8 $\pm$ 1.2 & 43.0 $\pm$ 1.0 & 27.2 $\pm$ 1.0 & 16.1 $\pm$ 0.7\\
RX J0437.2+3108 & K4 & 11.3 $\pm$ 0.8 & 82 $\pm$ 9 & -0.32 $\pm$ 0.05 & -0.21 $\pm$ 0.04 & 22.4 $\pm$ 1.6 & \nodata & \nodata & \nodata & \nodata\\
RX J0438.2+2023 & K2 & 16.1 $\pm$ 1.7 & Absorp. & 0.24 $\pm$ 0.06 & -0.13 $\pm$ 0.05 & 24.0 $\pm$ 1.7 & 25.3 $\pm$ 0.7 & 16.6 $\pm$ 0.7 & 10.5 $\pm$ 0.8 & 6.1 $\pm$ 0.5\\
RX J0438.2+2302 & M1 & 4.5 $\pm$ 0.4 & 111 $\pm$ 29 & -2.2 $\pm$ 1.0 & -0.3 $\pm$ 0.07 & 18.8 $\pm$ 1.4 & 18.2 $\pm$ 0.7 & 12.2 $\pm$ 0.7 & 7.6 $\pm$ 0.8 & 4.2 $\pm$ 0.5\\
HD 285957 & K2 & 22.5 $\pm$ 1.2 & Absorp. & -0.132 $\pm$ 0.014 & -0.15 $\pm$ 0.03 & 33 $\pm$ 2 & \nodata & \nodata & \nodata & \nodata\\
LkCa 15 & K5 & 13.9 $\pm$ 1.2 & 451 $\pm$ 51 & -15 $\pm$ 3 & -0.31 $\pm$ 0.06 & 24.2 $\pm$ 1.5 & 122.8 $\pm$ 1.5 & 94.7 $\pm$ 1.4 & 66.7 $\pm$ 1.4 & 69.3 $\pm$ 1.2\\
CoKu Tau4 & M1 & 25.8 $\pm$ 0.4 & 185 $\pm$ 33 & -1.8 $\pm$ 0.5 & -0.13 $\pm$ 0.04 & 23.8 $\pm$ 1.4 & 55.7 $\pm$ 1.0 & 37.1 $\pm$ 0.8 & 26.0 $\pm$ 0.8 & 16.5 $\pm$ 0.7\\
HD 283798 & G2 & 25.2 $\pm$ 1.2 & Absorp. & 1.96 $\pm$ 0.09 & -0.091 $\pm$ 0.018 & 62 $\pm$ 4 & \nodata & \nodata & \nodata & \nodata\\
RX J0444.4+1952 & M1 & 4.5 $\pm$ 0.4 & Absorp. & 0.09 $\pm$ 0.04 & -0.06 $\pm$ 0.02 & 20.5 $\pm$ 1.4 & \nodata & \nodata & \nodata & \nodata\\
HD 30171 & G5 & 108 $\pm$ 4 & Absorp. & 1.4 $\pm$ 0.2 & \nodata & 65 $\pm$ 4 & \nodata & \nodata & \nodata & \nodata\\
RX J0446.8+2255 & M1 & 8.0 $\pm$ 0.3 & 85 $\pm$ 4 & -1.3 $\pm$ 0.4 & -0.26 $\pm$ 0.04 & 20.9 $\pm$ 1.5 & \nodata & \nodata & \nodata & \nodata\\
RX J0447.9+2755 e & G2\tablenotemark{d} & 27.9 $\pm$ 1.4 & Absorp. & 1.2 $\pm$ 0.5 & \nodata & 19.8 $\pm$ 1.7 & \nodata & \nodata & \nodata & \nodata\\
RX J0447.9+2755 w & G2.5\tablenotemark{d} & 30.5 $\pm$ 1.8 & Absorp. & 1.035 $\pm$ 0.019 & -0.05 $\pm$ 0.02 & 20 $\pm$ 2 & \nodata & \nodata & \nodata & \nodata\\
UY Aur A+B a & K7 & 23.8 $\pm$ 1.3 & 324 $\pm$ 19 & -40 $\pm$ 11 & -0.8 $\pm$ 0.9 & 23.5 $\pm$ 1.5 & \nodata & \nodata & \nodata & \nodata\\
RX J0452.5+1730 & K4 & 8.8 $\pm$ 0.6 & 89 & -0.2 & -0.22 $\pm$ 0.04 & 22.2 $\pm$ 1.6 & 27.9 $\pm$ 0.8 & 16.8 $\pm$ 0.7 & 11.2 $\pm$ 0.8 & 7.3 $\pm$ 0.5\\
RX J0452.8+1621 & K6 & 24.9 $\pm$ 1.2 & 123 $\pm$ 12 & -0.7 $\pm$ 0.05 & -0.18 $\pm$ 0.04 & 26.3 $\pm$ 1.8 & 70.1 $\pm$ 1.2 & 44.4 $\pm$ 1.0 & 30.2 $\pm$ 1.0 & 17.2 $\pm$ 0.7\\
RX J0452.9+1920 & K5 & 4.8 $\pm$ 1.3 & 89 $\pm$ 8 & -0.43 $\pm$ 0.04 & -0.23 $\pm$ 0.02 & 26.9 $\pm$ 1.8 & \nodata & \nodata & \nodata & \nodata\\
HD 31281 & G0 & 79 $\pm$ 4 & Absorp. & 1.88 $\pm$ 0.09 & \nodata & 62 $\pm$ 4 & \nodata & \nodata & \nodata & \nodata\\
GM Aur & K7 & 14.8 $\pm$ 0.9 & 505 $\pm$ 11 & -88 $\pm$ 15 & -0.37 $\pm$ 0.08 & 22.1 $\pm$ 1.6 & 80.4 $\pm$ 1.2 & 56.4 $\pm$ 1.0 & 44.3 $\pm$ 1.0 & 47.8 $\pm$ 1.0\\
LkCa 19 & K0 & 20.1 $\pm$ 1.1 & 154 $\pm$ 40 & -0.9 $\pm$ 0.3 & -0.27 $\pm$ 0.03 & 37 $\pm$ 2 & 76.3 $\pm$ 1.2 & 47.9 $\pm$ 1.0 & 31.8 $\pm$ 0.8 & 18.4 $\pm$ 0.7\\
RX J0455.7+1742 & K3 & 12 $\pm$ 2 & Absorp. & 0.1 $\pm$ 0.04 & -0.131 $\pm$ 0.014 & 30 $\pm$ 2 & \nodata & \nodata & \nodata & \nodata\\
SU Aur & G2 & 59 $\pm$ 2 & 561 $\pm$ 58 & -5.0 $\pm$ 1.9 & 0.042 $\pm$ 0.012 & 78 $\pm$ 5 & 873 $\pm$ 5 & 816 $\pm$ 5 & 771 $\pm$ 5 & 976 $\pm$ 5\\
RX J0456.2+1554 & K7 & 9.7 $\pm$ 0.6 & 106 $\pm$ 20 & -0.7 $\pm$ 0.2 & -0.26 $\pm$ 0.05 & 25.0 $\pm$ 1.8 & \nodata & \nodata & \nodata & \nodata\\
HD 286179 & G0 & 17.1 $\pm$ 1.2 & Absorp. & 2.12 $\pm$ 0.09 & \nodata & 40 $\pm$ 2 & \nodata & \nodata & \nodata & \nodata\\
RX J0457.0+3142 & K2 & 25.5 $\pm$ 1.5 & Absorp. & 1.0 $\pm$ 0.5 & \nodata & 55 $\pm$ 3 & \nodata & \nodata & \nodata & \nodata\\
RX J0457.2+1524 & K1 & 42 $\pm$ 2 & Absorp. & 0.3 $\pm$ 0.07 & \nodata & 40 $\pm$ 2 & 110.8 $\pm$ 1.5 & 68.6 $\pm$ 1.2 & 47.5 $\pm$ 1.0 & 26.5 $\pm$ 0.9\\
RX J0457.5+2014 & K3 & 33 $\pm$ 3 & Absorp. & 0.39 $\pm$ 0.18 & \nodata & 30.8 $\pm$ 1.9 & 48.6 $\pm$ 1.0 & 30.0 $\pm$ 0.8 & 19.7 $\pm$ 0.8 & 11.6 $\pm$ 0.5\\
RX J0458.7+2046 & K7 & 7.8 $\pm$ 0.5 & Absorp. & 0.066 $\pm$ 0.017 & -0.167 $\pm$ 0.018 & 27.6 $\pm$ 1.7 & 42.0 $\pm$ 1.0 & 25.6 $\pm$ 0.8 & 17.3 $\pm$ 0.8 & 10.1 $\pm$ 0.5\\
RX J0459.7+1430 & K4 & 14.5 $\pm$ 0.6 & 53 & \nodata & -0.193 $\pm$ 0.017 & 24.0 $\pm$ 1.7 & 37.1 $\pm$ 0.8 & 23.0 $\pm$ 0.7 & 16.1 $\pm$ 0.8 & 8.8 $\pm$ 0.5\\
V836 Tau & K7 & 13.4 $\pm$ 1.1 & 403 $\pm$ 58 & -55 $\pm$ 17 & -0.36 $\pm$ 0.13 & 20.4 $\pm$ 1.4 & 72.9 $\pm$ 1.2 & 61.8 $\pm$ 1.2 & 52.3 $\pm$ 1.2 & 57.9 $\pm$ 1.0\\
RX J05072+2437 & K6 & 19.7 $\pm$ 1.0 & 126 $\pm$ 14 & -1.4 $\pm$ 0.4 & -0.3 $\pm$ 0.03 & 21.3 $\pm$ 1.4 & 26.0 $\pm$ 0.7 & 17.3 $\pm$ 0.7 & 12.7 $\pm$ 0.7 & 0.7 $\pm$ 0.5\\